\documentclass[aps, pre, twocolumn, showkeys, showpacs,amsmath,amssymb]{revtex4-1}
\usepackage{graphicx}
\usepackage{dcolumn}
\usepackage{bm}
\usepackage{transparent}
\usepackage{color}
\usepackage{savesym}
\savesymbol{a}
\savesymbol{b}
\savesymbol{d}
\savesymbol{e}
\savesymbol{k}
\savesymbol{l}
\savesymbol{L}
\savesymbol{p}
\savesymbol{r}
\savesymbol{s}
\savesymbol{S}
\savesymbol{T}
\savesymbol{F}
\savesymbol{R}
\savesymbol{H}
\savesymbol{t}
\savesymbol{w}
\savesymbol{th}
\savesymbol{div}
\savesymbol{tr}
\savesymbol{Re}
\savesymbol{Im}

\newcommand{\del}{\partial}

\newcommand{\G}{\Gamma}
\newcommand{\l}{\lambda}

\newcommand{\ph}{\phi}

\newcommand{\s}{\sigma}
\newcommand{\t}{\tau}
\newcommand{\th}{\theta}
\newcommand{\w}{\omega}

\newcommand{\dt}{{\Delta t}}
\newcommand{\dr}{{\Delta r}}
\newcommand{\dth}{{\Delta \theta}}
\newcommand{\dph}{{\Delta \phi}}
\newcommand{\dz}{{\Delta z}}

\newcommand{\De}{{\mathcal{D}e}}
\newcommand{\Re}{{\mathcal{R}e}}
\newcommand{\R}{{\mathcal{R}}}
\newcommand{\H}{{\mathcal{H}}}

\newcommand{\pdiff}[2]{\frac{\partial#1}{\partial #2}}

\graphicspath{{}}

\begin{document}
\title{Dean Instability in Double-Curved Channels}

\author{J.-D. Debus} \email{debusj@ethz.ch} \affiliation{ ETH
  Z\"urich, Computational Physics for Engineering Materials, Institute
  for Building Materials, Wolfgang-Pauli-Str. 27, HIT, CH-8093 Z\"urich
  (Switzerland)}
  
\author{M. Mendoza} \email{mmendoza@ethz.ch} \affiliation{ ETH
  Z\"urich, Computational Physics for Engineering Materials, Institute
  for Building Materials, Wolfgang-Pauli-Str. 27, HIT, CH-8093 Z\"urich
  (Switzerland)}

\author{H. J. Herrmann}\email{hjherrmann@ethz.ch} \affiliation{ ETH
  Z\"urich, Computational Physics for Engineering Materials, Institute
  for Building Materials, Wolfgang-Pauli-Str. 27, HIT, CH-8093 Z\"urich
  (Switzerland)}

\begin{abstract}
We study the Dean instability in curved channels using the lattice Boltzmann model for generalized metrics. For this purpose, we first improve and validate the method by measuring the critical Dean number at the transition from laminar to vortex flow for a streamwise curved rectangular channel, obtaining very good agreement with the literature values. Taking advantage of the easy implementation of arbitrary metrics within our model, we study the fluid flow through a double-curved channel, using ellipsoidal coordinates,  and study the transition to vortex flow in dependence of the two perpendicular curvature radii of the channel. We observe not only transitions to 2-cell vortex flow, but also to 4-cell and even 6-cell vortex flow, and we find that the critical Dean number at the transition to 2-cell vortex flow exhibits a minimum when the two curvature radii are approximately equal.
\end{abstract}

\maketitle

\section{Introduction}\label{sec:introduction}

Curved channel flow has attracted fluid dynamics researchers for over a
hundred years because of its practical applications in engineering and fascinating features at high Reynolds numbers. At low Reynolds numbers, the flow through a curved duct is laminar, and its velocity profile resembles the parabolic profile of plane channel flow. Following Ref. \cite{finlay1988}, we will refer to this type of flow as ``curved channel Poiseuille flow''. For increasing Reynolds number, the flow pattern is determined by centrifugal forces until it becomes unstable and bifurcates to secondary flow. 
 In 1928, W. R. Dean showed analytically for the case of a
narrow channel that the fully developed flow between two concentric
cylinders (see Fig. \ref{fig:cylinder_geometry}) can be characterized by a single non-dimensional parameter, the Dean
number $\De$, which depends on the Reynolds number $\Re$ as well as on
the channel geometry \cite{dean1928}. It is defined as 
\begin{align*}
	\De := \Re \cdot \sqrt{\frac{d}{\R}},
\end{align*}
where $d = r_o - r_i$ is the channel width, $r_i, r_o$ are
the radii of the inner and outer cylinder respectively and $\R$ denotes the
curvature radius of the inner wall (for a cylinder, $\R = r_i$). In
particular, Dean performed a stability analysis to show that flow between two concentric cylinders becomes unstable for
small perturbations when the Dean number $\De$ exceeds a critical value of
$\De_c \approx 36$.  Above this value, pairs of counter-rotating
streamwise-oriented vortices develop, which is known as Dean vortex flow.  
The theoretical predictions by Dean have been confirmed in various
experiments, see for example H\"ammerlin 
\cite{hammerlin1957} or Brewster, Grosberg \& Nissan \cite{brewster1959}. 
Numerical simulations of curved channel flow have also been popular, see for example Finlay \textit{et al.}
\cite{finlay1988, finlay1989, finlay1990}, who extended Dean's 
stability criterion to channels of different aspect ratios $\eta =
r_i/r_o$, using high-accuracy linear stability analysis. 

All the studies mentioned above deal with a very idealized channel geometry, namely with channels having a rectangular cross section, which are uniformly curved along the streamwise direction only. Thinking of practical engineering applications (e.g. water ducts), however, channels do not possess a perfectly rectangular cross section, but can also be bent perpendicularly to the streamwise direction for instance under the influence of external forces such as gravity. The resulting cross section becomes the section of a circular ring, such that the channel possesses a second, cross-sectional curvature. In this paper, we study the Dean instability in such a double-curved channel, which, to the best of our knowledge, has never been done before. Introducing a second curvature, we expect qualitative deviations from the behaviour of the Dean flow as compared to the idealized cylindrical case, since the translational symmetry along the cylinder axis is then broken. 

For the double-curved channel, we study the flow at different Dean numbers and observe a bifurcation from curved channel Poiseuille flow to vortex flow, which is similar to the vortex flow in the cylindrical channel, showing a pair of counter-rotating vortices. We measure the critical Dean number at the bifurcation point in dependence of the two perpendicular curvatures. 

First, we vary the streamwise curvature radius $\R_\th$ while keeping the cross-sectional curvature radius $\R_\ph$ fixed. Surprisingly, we find that the lowest Dean number at which the Dean instability occurs corresponds to the configuration in which both curvature radii are approximately equal, $\R_\th \approx \R_\ph$. For channels with both weaker or stronger streamwise curvature, we measure an (almost linear) increase of the critical Dean number with $\R_\th$. For strongly-curved channels, we even observe a second bifurcation to vortex flow with 4 counter-rotating vortices.

Secondly, we vary the cross-sectional curvature radius $\R_\ph$ while keeping the streamwise curvature fixed. Again, we find a minimum of the critical Dean number for the case $\R_\th \approx \R_\ph$, consistent with the first study. For increasing $\R_\ph$, the curvature perpendicular to the streamwise direction becomes smaller and smaller, such that the double-curved channel approaches a cylindrical geometry in the limit $\R_\ph \rightarrow \infty$. Correspondingly, for increasing $\R_\ph$, the critical Dean number for the double-curved channel approaches the value of the critical Dean number for the cylindrical channel in our simulations. 

\medskip

For the simulations, we use the lattice Boltzmann (LB) method, which has been developed during the last decades to simulate fluids by means of simple arithmetic
operations instead of discretizing and solving the complicated macroscopic
equations of continuum fluid mechanics. The method itself is based on
the Boltzmann kinetic equation, which describes the motion of the microscopic
fluid particles instead of the macrosopic continuum. Because of its simplicity
and straightforward parallelizability, the LB method has gained
more and more popularity among scientists and engineers in the past. A review
about the LB method is given in Ref. \cite{chen1998}. While it was originally designed to solve fluid flows, the LB method has even been applied
to electrodynamics \cite{mendoza2010electro} and magnetohydrodynamics
\cite{mendoza2008} as well as relativistic \cite{mendoza2010rel} and
ultra-relativistic flows \cite{mohseni2013}. Most of the LB applications use standard Cartesian coordinates (e.g. for
fluid flow in rectangular cavities), which cannot be applied
straightforwardly to more complex curved geometries. However, the
lattice Boltzmann method has recently been extended to general metrics being
defined by a metric tensor \cite{mendoza2013, mendoza2014}, which
offers a variety of interesting new applications.
With this extension at hand, it becomes possible to simulate fluids in
\textit{arbitrary} geometries, while standard LB methods are
restricted to simple geometries. 

In this paper, we improve the method of Ref. \cite{mendoza2013} by increasing the accuracy of the forcing term in the LB equation. The improvement is validated for the case of a cylindrical channel, for which we measure the dependence of the critical Dean number on the channel aspect ratio $\eta =r_i/r_o$. Comparing our results to the numerical values given in Ref. \cite{finlay1990} by
Finlay \textit{et al.}, we find very good agreement.

Since our model can handle arbitrary geometries, we can easily introduce a second perpendicular curvature in the channel by choosing ellipsoidal coordinates. To this end, we only have to adapt the metric tensor, which is much simpler then deriving a new expression for the LB equation
for each special choice of coordinates, as it is
commonly done in the literature for very simple geometries (see e.g. Ref.
\cite{zhou2008}).

\section{Lattice Boltzmann Method}\label{sec:method}

In this section, we will shortly review the method used to simulate the motion 
of the fluid. For further details
we refer to \cite{mendoza2013,mendoza2014}. 

The method is based on the Boltzmann equation, which describes the motion of
 fluid particles in terms of a distribution function $f$. In a curved 
three-dimensional Riemann space, the Boltzmann equation reads
\begin{align}\label{eq:B}
	\pdiff{f}{t} + \xi^i \pdiff{f}{x^i} + F^i
	\pdiff{f}{\xi^i} = \mathcal C[f]
\end{align}
\cite{mendoza2013}, where $f = f(x^i, p^i)$ denotes the distribution
function, which depends on the space coordinates $x^i = (x^1,x^2,x^3)$ as well as on the momentum $p^i = m (\xi^1,\xi^2,\xi^3)$, where $m$ and $\xi^i$ denote the mass and velocity of the fluid particles respectively, and the mass is set to $m=1$. We are using the Einstein sum convention throughout the whole paper, i.e. Latin indices run over the spatial directions 1 to 3. The spatial metric enters the Boltzmann equation through the force $F^i := - \Gamma^i_{jk} \xi^j \xi^k$, which depends on the Christoffel symbols $\Gamma^i_{jk}$ and thus drives the particles along the geodesics of the curved space. Collisions between fluid particles are accounted for by the collision operator $\mathcal C[f]$, for which we use the
Bhatnagar-Gross-Krook (BGK) approximation \cite{BGK}
\begin{align*}
	\mathcal C[f] = - \frac{f - f^{\rm eq}}{\t},
\end{align*}
where $\t$ denotes the relaxation parameter and $f^{\rm eq}$ is the
Maxwell-Boltzmann equilibrium distribution. The latter is given by 
\begin{align*}
	f^{\rm eq} = \frac{\rho}{ \left(2\pi \theta_T\right)^{3/2}} \exp \left[-
	\frac{1}{2\theta_T}  \left(\xi^i - u^i\right) g_{ij} \left(\xi^j -
	u^j\right)\right]
\end{align*}
\cite{mendoza2013}, where $\theta_T$ is the normalized temperature, $\rho$ and 
$u^i$ denote the macroscopic density and velocity of the fluid 
(normalized by the speed of sound $c_s$)
and $g_{ij}$ are the components of the metric tensor $g$. In the following, we will always assume the isothermal case $\theta_T = 1$. 

The macroscopic density $\rho$ and velocity $\vec u$ are given by the zeroth and
first order moment of the distribution function,
\begin{align}
	\label{eq:moment}
	\rho = \int f \,\sqrt g \, d^3 \xi, \qquad
	\rho u^i = \int f \,\xi^i \sqrt g \, d^3 \xi,
\end{align}
where $\sqrt g \, d^3 \xi:= \sqrt{\det g} \, d\xi^1 d\xi^2 d\xi^3$ denotes the
invariant volume element of the momentum space. Since conservation of mass and momentum are
intrinsic features of the Boltzmann equation, the macroscopic density $\rho$ and
velocity $u^i$ automatically fulfil the hydrodynamic conservation equations,
which can be shown rigorously by a Chapman-Enskog expansion \cite{chapman1970}.  
In covariant form, the macroscopic conservation equations read
 \begin{align*} 
	\pdiff{\rho}{t} + \nabla_i \left(\rho u^i \right) = 0, \qquad
	\pdiff{}{t} \left(\rho u^i \right) + \nabla_j T^{ij} = 0,
\end{align*} 
where $\nabla$ denotes the covariant derivative (Levi-Civita connection) and
$T^{ij}$ is the energy-stress tensor. Explicitly, the energy-stress tensor 
is given by $T^{ij} = P g^{ij} + \rho u^i u^j - \mu (g^{lj} \nabla_l u^i +
g^{il} \nabla_l u^j)$, where $P = \rho \theta$ is the hydrostatic pressure, $\mu$ is the dynamic shear viscosity 
and $g^{ij}$ denote the components of the inverse metric tensor. 

\medskip

In order to obtain the lattice version of the Boltzmann equation, the coordinate and momentum space are discretized on a (sufficiently symmetric) 
lattice, and all vectors are expressed in terms of a commuting basis 
$(\mathbf e_1, \mathbf e_2, \mathbf e_3) = (\frac{\del}{\del x^1},
\frac{\del}{\del x^2}, \frac{\del}{\del x^3})$, which defines the coordinate
frame on the lattice.
The minimum configuration of discrete velocities to fulfil all necessary 
symmetry relations is given by the D3Q41 lattice proposed in Ref. \cite{chikatamarla2009}. 
This lattice contains a set of 41 discrete velocities
$\{\vec c_\l\}_{\l=1}^{41}$ (see Table \ref{tab:weights}), which are normalized by the speed of sound $c_s$, i.e. $\vec \xi_\l = \vec c_\l/c_s $. The value of the
speed of sound for this specific lattice is given by $c_s^2 =
1-\sqrt{2/5}$. From the discretized distribution function $f_\l(\vec x,
t) := f(\vec x, \vec \xi_\l, t)$, the macroscopic quantities
of the fluid, i.e. the density $\rho$ and the macroscopic velocity $\vec u$, are recovered by taking moments of the distribution function,
\begin{align*}
	\sum_{\l=1}^{41} f_\l = \rho' 
	\quad,\quad
	\sum_{\l=1}^{41} c_\l^i f_\l = \rho' u^i,
\end{align*}
where $\rho' = \rho / \sqrt g$ (the factor $\sqrt g$ in the invariant
integration measure in Eq. (\ref{eq:moment}) is
absorbed into a redefinition of the fluid density).

% ---------------------------------------
\begin{table}
  \centering\setlength\extrarowheight{4pt}
  \begin{tabular}{|@{\quad}c@{\quad}|@{\quad}c@{\quad}|@{\quad}c@{\quad}|}\hline
    $\l$ & $\vec c_\l$ & $w_\l$ \\[4pt] \hline\hline
    1 & $(0,0,0)$ & $\frac{2}{2025} \left(5045-1507\sqrt{10}\right)$ \\[4pt] \hline
    2,3 & $(\pm 1,0,0)$ &  \\ 
    4,5 & $(0,\pm 1,0)$ & $\frac{37}{5\sqrt{10}} - \frac{91}{40}$ \\ 
    6,7 & $(0,0,\pm 1)$ &  \\[4pt] \hline
    8-11 & $(\pm 1,\pm 1, 0)$ &  \\ 
    12-15 & $(\pm 1,0,\pm 1)$ & $\frac{1}{50} 	\left(55-17\sqrt{10}\right)$ \\ 
    16-19 & $(0,\pm 1,\pm 1)$ &  \\[4pt] \hline
    20-27 & $(\pm 1,\pm 1,\pm 1)$ & $\frac{1}{1600}
    \left(233\sqrt{10}-730\right)$ \\[4pt] \hline 
    28,29 & $(\pm 3,0,0)$ &  \\ 
    30,31 & $(0,\pm 3,0)$ & $\frac{1}{16200} \left(295-92\sqrt{10}\right)$ \\ 
    32,33 & $(0,0,\pm 3)$ &  \\[4pt] \hline
    34-41 & $(\pm 3,\pm 3,\pm 3)$ & $\frac{1}{129600}
    \left(130-41\sqrt{10}\right)$ \\[4pt] \hline
  \end{tabular}
  \caption{Discrete velocity vectors $c_\l$ of the D3Q41 lattice and the
  corresponding weights $w_\l$ in the Hermite expansion.} 
  \label{tab:weights}  
\end{table}
% ---------------------------------------

In terms of discretized quantities, the Boltzmann equation (\ref{eq:B}) becomes
\begin{align}
	\label{eq:LB}
	f_\l(x^i + c_\l^i \dt, t + \dt) - f_\l(x^i, t) 
	 = - \frac{\dt}{\t} \left( f_\l - f_\l^{\rm eq} \right) + \dt
	 \mathcal F_\l,
\end{align}
where the left-hand side represents free streaming, whereas the right-hand side
corresponds to particle collisions. The forcing term $\mathcal F_\l$ contains
the information about the geometry of the space, encoded in a combination of Christoffel symbols. 
The relaxation parameter $\tau$ is directly related to the dynamic shear viscosity $\mu$ 
as well as to the kinematic viscosity $\nu$ by
\begin{align}\label{eq:viscosity}
	\mu = \rho \nu = \rho \left(\tau - \frac{1}{2} \right) c_s^2 \dt.
\end{align}

In order to obtain discrete expressions for the equilibrium distribution
$f_\l^{\rm eq}$ and for the forcing term $\mathcal F_\l$, we expand the
distribution function into tensor Hermite polynomials, defined by
\begin{align*}
	\H_{(n)}^{i_1,\ldots,i_n}(\vec \xi) = (-1)^n w(\vec \xi)^{-1} \frac{\del}{\del\xi_{i_1}} \cdots \frac{\del}{\del\xi_{i_n}} w(\vec \xi),
\end{align*}
where $w(\vec \xi)$, the weight function, is given by
\begin{align*}
	w(\vec\xi) = \frac{1}{(2\pi)^{3/2}} \exp \left(- \frac{1}{2}|\vec \xi|^2
	\right).
\end{align*} 
This yields
\begin{align}\label{eq:hermite-expansion}
	f(\vec x, \vec \xi, t) = w(\vec \xi) \sum_{n=0}^\infty \frac{1}{n!}
	a_{(n)}^{i_1,\ldots,i_n}(\vec x,t) \,\H_{(n)}^{i_1,\ldots,i_n}(\vec \xi),
\end{align}
where $a_{(n)}$ are the coefficients of the expansion.
For the equilibrium distribution, $f = f^{\rm eq}$, the first four expansion
coefficients are given by
\begin{align}
	\nonumber
	&a_{(0)}^{{\rm eq}} = \rho',\qquad
	a_{(1)}^{{\rm eq},i} = \rho' u^i,\qquad 
	a_{(2)}^{{\rm eq},ij} = \rho' c_s^2 \Delta^{ij} + \rho' u^i u^j, \\
	\label{eq:hermite-coefficients-eq}
	&a_{(3)}^{{\rm eq},ijk} = \rho' c_s^2 \left( \Delta^{ij} u^k + \Delta^{jk} u^i +
	\Delta^{ki} u^j \right) + \rho' u^i u^j u^k,
\end{align}
where 
$\Delta^{ij} := g^{ij} - \delta^{ij}$ is a measure for the deviation from
flat space and factors $c_s^2$ account for a normalization of the velocities $\vec \xi
\rightarrow \vec c_\l / c_s$. Now, we obtain the lattice version of the equilibrium
distribution by applying the Gauss-Hermite quadrature rule, which in this case is equivalent to replacing $w(\vec \xi)$ by $w_\l$, where the discrete weights $w_\l$ are given in Table \ref{tab:weights}. For the D3Q41 lattice, Gauss-Hermite quadrature preserves the first four moments of the distribution function \textit{exactly}. Therewith, the lattice equilibrium function is given by
\begin{align*}
	f_\l^{\rm eq} &= w_\l \bigg( a_{(0)}^{{\rm eq}} 
	+ \frac{1}{c_s^2} a_{(1)}^{{\rm eq},i} c_\l^i
	+ \frac{1}{2! c_s^4} a_{(2)}^{{\rm eq},ij} \Big( c_\l^i c_\l^j - c_s^2 \delta^{ij} \Big)\\
	&+ \frac{1}{3! c_s^6} a_{(3)}^{{\rm eq},ijk} \Big( c_\l^i c_\l^j
	c_\l^k - c_s^2 \big( \delta^{ij} c_\l^k + \delta^{jk}
	c_\l^i +   \delta^{ki} c_\l^j \big) \Big) \bigg).
\end{align*}

In order to calculate the forcing term $\mathcal F_\l$, we rewrite 
\begin{align}\label{eq:forcing}
	F^i	\pdiff{f}{\xi^i} = - w \sum_{n=0}^\infty \frac{1}{n!}
	a_{(n)}^{i_1,\ldots,i_n} \, F^i \,\H_{(n+1)}^{i, i_1,\ldots,i_n},
\end{align}
where we have expressed $f$ in terms of its Hermite expansion (\ref{eq:hermite-expansion}). As an improvement of the forcing term in Ref. \cite{mendoza2014}, we do not
approximate all the expansion coefficients $a_{(n)}$ by the coefficients of the 
equilibrium distribution (\ref{eq:hermite-coefficients-eq}), but set
\begin{align*}
	&a_{(0)} = a_{(0)}^{{\rm eq}}, \qquad
	a_{(1)}^{i} = a_{(1)}^{{\rm eq},i},
	&a_{(2)}^{ij} = a_{(2)}^{{\rm eq},ij} - \s^{ij},
\end{align*}
where $\s^{ij} = -( 1 - \frac{1}{2\t}) \sum_\l c_\l^i c_\l^j \left( f_\l - f^{{\rm eq}}_\l \right)$ denotes the stress tensor, and the factor $(1 - \frac{1}{2\t})$ accounts for discrete lattice effects.
Finally, we obtain the discrete forcing term by plugging these coefficients into
Eq. (\ref{eq:forcing}) (using normalized velocities $\vec c_\l$),  and truncating the expansion at third order:
\begin{align*}
	\mathcal F_\l &= w_\l \bigg( 
	\frac{1}{c_s^2} a_{(0)} F_\l^i c_\l^i
	+ \frac{1}{c_s^4} a_{(1)}^{i} F_\l^j \Big( c_\l^i c_\l^j - c_s^2 \delta^{ij} \Big)\\
	&+ \frac{1}{2 c_s^6} a_{(2)}^{ij} F_\l^k \Big( c_\l^i c_\l^j
	c_\l^k - c_s^2 \big( \delta^{ij} c_\l^k + \delta^{jk}
	c_\l^i +   \delta^{ki} c_\l^j \big) \Big) \bigg).
\end{align*}
where $F_\l^i = - \Gamma^i_{jk} c_\l^j c_\l^k$. Additional external forces can
be added straightforwardly by replacing $F_\l^i \rightarrow F_\l^i + F_{\rm
ext}^i$.

Having all ingredients at hand for the LB equation
(\ref{eq:LB}), the LB algorithm can be applied as usual: 
After assigning initial conditions to the macroscopic quantities $\rho$ and
$\vec u$, the distribution function $f$ is successively updated time step by 
time step according to the LB equation. This is done by
dividing each time step into a free streaming step (corresponding to the 
left-hand side of Eq. (\ref{eq:LB})) and a collision step (corresponding to the
right-hand side of Eq. (\ref{eq:LB})). Special care has to be taken for the
boundary conditions, which will be addressed in the next sections in the
context of concrete examples.

\section{Validation: Flow through Cylindrical Channel}\label{sec:cylinder}

\begin{figure}
	\includegraphics[width=.8\columnwidth]{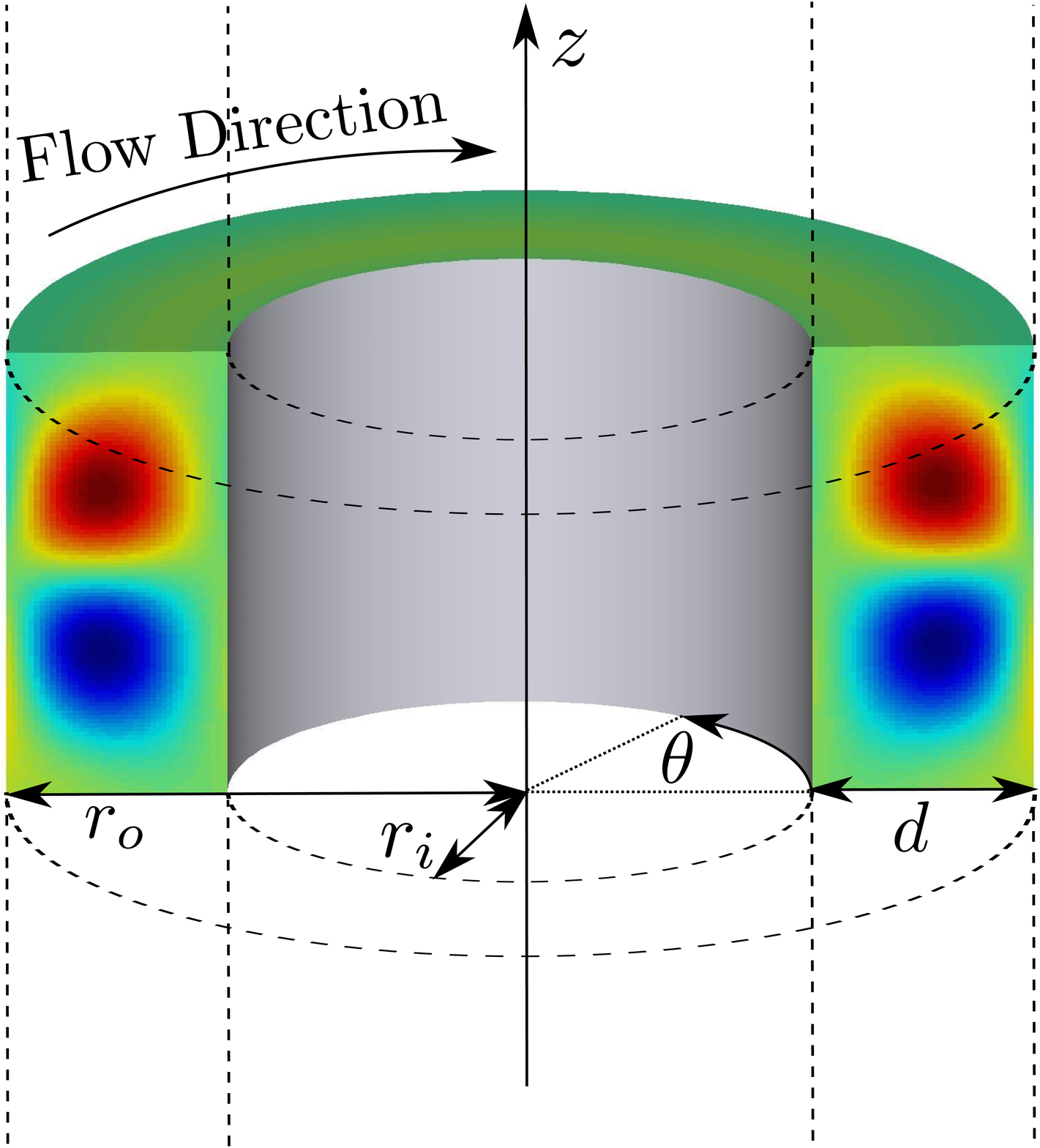}
	\caption{(Color online) Geometry of the curved channel in cylindrical coordinates $(r,\th,z)$. The colored cross-sections depict the vorticity of the axisymmetric flow, where the blue (lower) and red (upper) spots correspond to low and high values of the vorticity, respectively. The dashed lines indicate the periodicity of the channel in the $\th$- and $z$-direction.}
	\label{fig:cylinder_geometry}
\end{figure}
For the validation of the improved model, we consider the idealized case of flow in a closed cylindrical channel, as depicted in Fig. \ref{fig:cylinder_geometry}. In an experimental application, the channel would not be completely closed, but would have a finite opening angle $< 2\pi$ with open boundaries at the inlet and outlet. However, for long and narrow channels, as considered in this study, the ratio between the streamwise extent and the spanwise extent $d$ (being relevant for the instability) is very small ($\sim 10^{-2}$) such that possible finite-size effects at the inlet/outlet can be neglected. In order to drive the fluid, a pressure gradient between the inlet and outlet of a channel can be applied. In our simulations, we use an external force in the azimuthal direction instead, which for incompressible flow is equivalent to a pressure gradient.

Depending on the aspect ratio of the channel as well as on the Reynolds number $\Re$, the primary Poiseuille flow is expected to undergo a transition to secondary flow (Dean flow) at a specific critical value of the Dean number,
\begin{align}\label{eq:dean-number}
	\De := \Re \cdot \sqrt{\frac{d}{\R}} 
	= \frac{\langle v_\theta \rangle d}{\nu} \cdot \sqrt{\frac{d}{\R}},
\end{align}
where $\langle v_\theta \rangle$ denotes the mean azimuthal fluid velocity, $\nu =
\left( \tau - 1/2 \right) c_s^2 \dt$ is the kinematic viscosity, $d = r_o -
r_i$ is the channel width, $r_i$ and $r_o$ denote the inner and outer radius, respectively, and $\R$ is the curvature radius of the inner wall
(for a cylinder $\R = r_i$).
The secondary flow occurs due to centrifugal instabilities and is characterized
by a pair of counter-rotating vortex tubes which are oriented along the stream direction.
Using the LB method as described in section \ref{sec:method}, we
have modeled the axisymmetric fluid flow for different channel aspect ratios 
$\eta = r_i/r_o$. In order to avoid staircase approximations at the channel boundaries, we use cylindrical coordinates $(x^1,x^2,x^3) = (r,\th,z)$, which are perfectly adapted to the geometry of the channel. In these coordinates, the metric tensor is given by
\begin{align*}
	g = \begin{pmatrix}
			1 & 0 & 0  \\
			0 & r^2 & 0 \\
			0 & 0 & 1
		\end{pmatrix},
\end{align*}
and we express all vector fields and tensors in terms of the standard basis
$(\mathbf e_r, \mathbf e_\th, \mathbf e_z) = (\frac{\del}{\del r}, \frac{\del}{\del \th},
\frac{\del}{\del z})$, for example $u = u^r \mathbf e_r + u^\th \mathbf e_\th +
u^z \mathbf e_z$.
Since this basis is commuting, Christoffel symbols can be
calculated by
\begin{align*}
	\Gamma^i_{jk} = \frac{1}{2} g^{im} \left(\pdiff{g_{mj}}{x^k} + \pdiff{g_{mk}}{x^j}
	- \pdiff{g_{jk}}{x^m} \right),
\end{align*}  
which yields the only non-vanishing Christoffel symbols $\Gamma^r_{\th \th} =
-r$, $\Gamma^\th_{r \th} = \Gamma^\th_{\th r} = 1/r$. 

Since the cylindrical channel is translational invariant in $\th$, we impose periodic boundaries in $\th$, which effectively reduces the
three-dimensional problem to a two-dimensional one. Since we are only interested in ``two-dimensional'' vortex solutions (as they are called in Ref. \cite{guo1991splitting}), each velocity component depends only on the spanwise directions $r$ and $z$. 
For the simulations, we use a rectangular lattice of size $L_r \times L_\th \times L_z = 128 \times 1
\times 256$, where each lattice node is labeled by integer lattice indices 
$(\hat r, \hat \th, \hat z) \in [1,L_r] \times [1,L_\th] \times [1,L_z]$. The
transformation between lattice units and physical units is given by
\begin{align*}
	r = r_i + \hat r \cdot \dr , \quad
	\th = \hat \th \cdot \dth , \quad
	z = \hat z \cdot \dz,
\end{align*}
where $\dr$, $\dth$ and $\dz$ denote the lattice spacings in the radial, azimuthal
and axial direction respectively. In our simulations, we set $\dr =
\dth = \dz = \dt$ and $r_i = 1$. Defining the aspect ratio as $\eta = r_i/r_o$, we
obtain for the lattice spacing $\dr = (1 - \eta)/(\eta L_r)$. In order to compare our results to the work by Finlay et al. \cite{finlay1990}, we only vary the radial aspect ratio $\eta = r_i/r_o$, keeping the spanwise aspect ratio fixed, $L_z/L_r = 2$. 
 At the two channel walls at $r =r_i$ and $r = r_o$, we impose Dirichlet boundary conditions on the fluid
velocity, $u(r_o) = u(r_i) = 0$. This condition enters the algorithm through
the equilibrium distribution $f^{\rm eq}$, which is evaluated with $u = 0$ at $r =
r_i$ and $r = r_o$ at each time step. In the $\th$- and $z$-directions, we use
periodic boundary conditions. We note that by considering periodicity in $z$, a symmetry condition is imposed on the solution, since the spanwise wave length of the Dean vortices is restricted to the values $\lambda = 2 L_z/k$ (where $k$ denotes the number of vortex cells). However, the choice $L_z/L_r = 2$ is motivated by the fact that the channel cross-section is perfectly adapted to the 2-cell vortex solution, since in this case the Dean cells can occupy the total section.

The relaxation parameter $\tau$ can be used to tune the Reynolds number (and thus the Dean number) to the parameter range of interest, where the bifurcation occurs. For the simulations, we choose a $\tau = 0.9$, which also enhances fast convergence to the stationary state.

In order to study the bifurcation from curved channel Poiseuille flow to secondary Dean flow, we
vary the Dean number $\De$ by varying the strength of the driving force. At the
critical Dean number $\De_c$, we expect the formation of counter-rotating vortex tubes, which should increase in strength for higher values of $\De$. This can be
measured by calculating the vorticity of the flow, given by $\vec \w =
\vec \nabla \times \vec u$. (Note that some authors use the helicity, $\vec h = \vec u \times \vec \w$, instead, to analyze the Dean vortices. However, for the streamwise oriented vortices considered in this study, both quantities are equivalent, since the vorticity is fully characterized by its streamwise component, $\w^\th$). In particular, we are interested in the mean absolute vorticity, averaged over the whole cross-section of the channel,
\begin{align*}
	\langle \w^\th \rangle = \frac 1 S \int_S \left| \pdiff{u^r}{z} -
	\pdiff{u^z}{r}\right| r\, dr dz,
\end{align*}
where $S = \int dr dz$ denotes the cross-sectional area. At the critical Dean
number $\De_c$, the vorticity is expected to increase considerably. Indeed, this
behaviour has been observed in the simulations. 

\begin{figure}
	\includegraphics[width=\columnwidth]{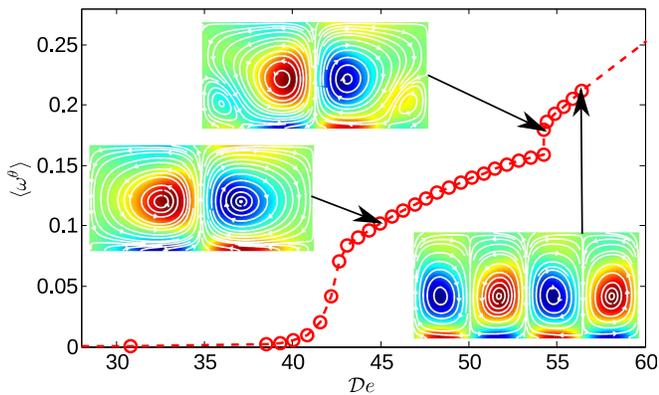}
	\caption{(Color online) Average vorticity $\langle \w^\th \rangle$ depending on the Dean number $\De$ for a cylindrical channel with aspect ratio $\eta = r_i/r_o = 0.80$. As can be seen, the bifurcation from Poiseuille flow to 2-cell vortex flow  occurs at $De_c \approx 42$, followed by a second bifurcation to 4 vortices  at $\De_{c2} \approx 54$.
	The colored pictures depict the velocity streamlines on a channel cross-section perpendicular to the stream direction. The colors represent the strength of the  vorticity, where blue and red colors correspond to clockwise and counterclockwise rotating vortices, respectively.}
	\label{fig:paper_cylinder_vorticity_plot}
\end{figure}

\begin{figure}
	\includegraphics[width=0.8\columnwidth]{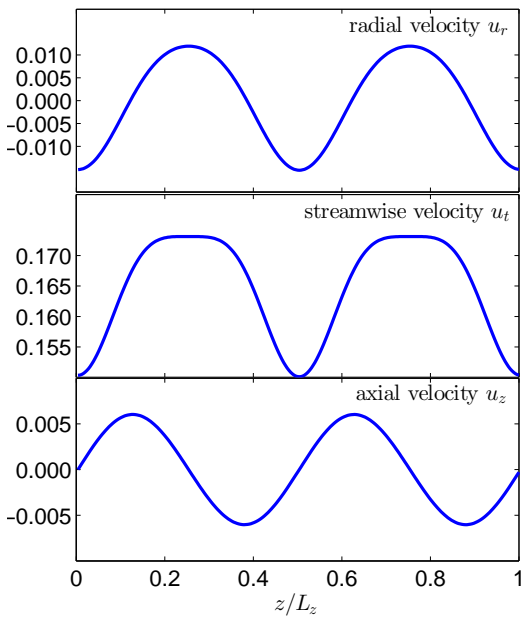}
	\caption{(Color online) Radial, streamwise and axial velocity profiles for the 4-cell vortex flow for a cylindrical channel with aspect ratio $\eta = 0.80$ at Dean number $\De = 56$}
	\label{fig:paper_cylinder_velocity_plot}
\end{figure}

Exemplarily, Fig. \ref{fig:paper_cylinder_vorticity_plot} shows the average vorticity $\langle \w^\th \rangle$ depending on the Dean number $\De$ for a channel with aspect ratio $\eta = 0.80$. The curve agrees with the expectations, showing a transition from zero-vorticity Poiseuille flow for $\De < 42$ to vortex flow for $\De > 42$. The behavior of the vorticity at $\De_c \approx 42$ is indicative of an imperfect bifurcation, as can be seen in the recent work by Haines \textit{et. al} in Ref. \cite{haines2013dean}. At $\De_c$, two counter-rotating vortex tubes form, which
increase in strength for higher Dean numbers. The colored pictures in Fig. \ref{fig:paper_cylinder_vorticity_plot} illustrate the velocity streamlines on a cross-section of the channel perpendicular to the stream direction at different Dean numbers. The colors represent the strength of the streamwise vorticity $\w^\th$, where blue and red colors correspond to clockwise and counterclockwise rotating vortices, respectively.
Fig. \ref{fig:paper_cylinder_vorticity_plot} also reveals a second bifurcation, which is indicated by a further increase of the vorticity at $\De_{c2} \approx 54$, where a second pair of vortex tubes begins to form. Fig. \ref{fig:paper_cylinder_velocity_plot} shows the radial, streamwise and axial velocity profiles versus the $z$-position for the 4-cell vortex flow for $\eta = 0.80$ and $\De = 56$.

\begin{figure}
	\includegraphics[width=\columnwidth]{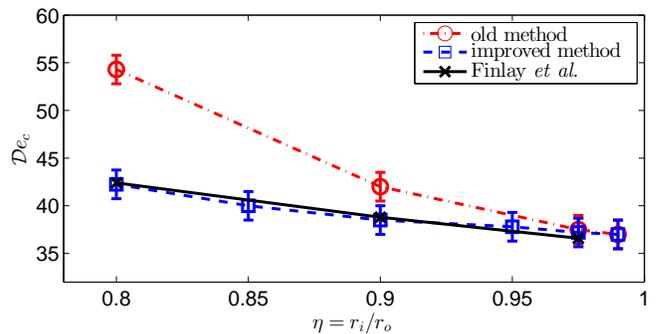}
	\caption{(Color online) Critical Dean number $\De_{c}$ at the bifurcation from Poiseuille flow to 2-cell vortex flow as function of the aspect ratio $\eta = r_i/r_o$ for a cylindrical channel. The results obtained by our improved method agree very well with the literature values of Finlay \textit{et al.}}
	\label{fig:critical_dean_1st}
\end{figure}

We have measured the critical Dean number at the bifurcation from Poiseuille flow to 2-cell vortex flow for different channel aspect ratios $\eta = r_i/r_o$. The results are shown in Fig. \ref{fig:critical_dean_1st}, which depicts the dependence of the critical Dean number on the aspect ratio of the channel. The errorbars result from the uncertainty in determining the critical Dean number from the vorticity curve, since the vorticity increases smoothly at the bifurcation point  (as can be seen in Fig. \ref {fig:paper_cylinder_vorticity_plot}). We have also compared the present improved version of our method with the old version used
in a previous publication \cite{mendoza2013}, where the moments of the
distribution function $f$ in the forcing term have been approximated by the 
moments of the equilibrium distribution function $f^{\rm eq}$. Comparing both
methods to the numerical results by Finlay \textit{et al.}
\cite{finlay1990} in Fig. \ref{fig:critical_dean_1st}, we observe that the present improved method agrees very well
with the literature, whereas our old method leads to deviations from the literature values. 
It can be seen that the deviations between the old method and the improved
method vanish for $\eta \rightarrow 1$. This means that the error in
the approximation of the forcing term in the old method becomes negligible when
the resolution is sufficiently high.

\begin{figure}
	\includegraphics[width=\columnwidth]{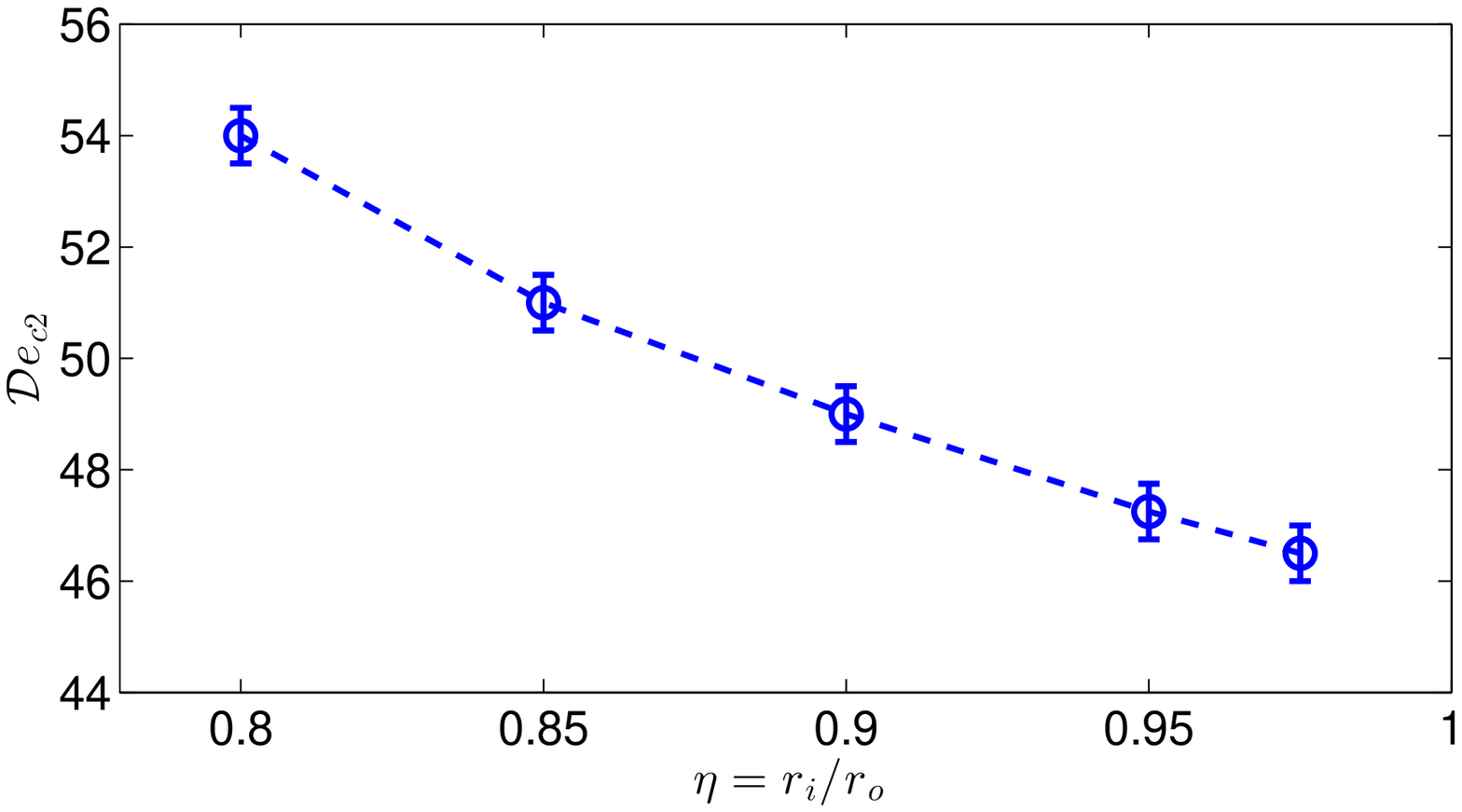}
	\caption{(Color online) Second critical Dean number $\De_{c2}$ at the bifurcation from 2-cell vortex flow to 4-cell vortex flow } as function of the aspect ratio $\eta = r_i/r_o$ for a cylindrical channel.
	\label{fig:critical_dean_2nd}
\end{figure}

We have also plotted the second critical Dean number $\De_{c2}$, at which the second bifurcation from 2-cell vortex flow to 4-cell vortex flow occurs. The dependence
on the aspect ratio $\eta$ is shown in Fig. \ref{fig:critical_dean_2nd}. As can be seen, $\De_{c2}$ decreases monotonically with the aspect ratio. For very narrow channels with aspect ratio $\eta > 0.95$, the second bifurcation occurs already at relatively low Dean numbers between $45$ and $50$.

\subsection*{Dependence on the Resolution}

\begin{figure}
	\includegraphics[width=\columnwidth]{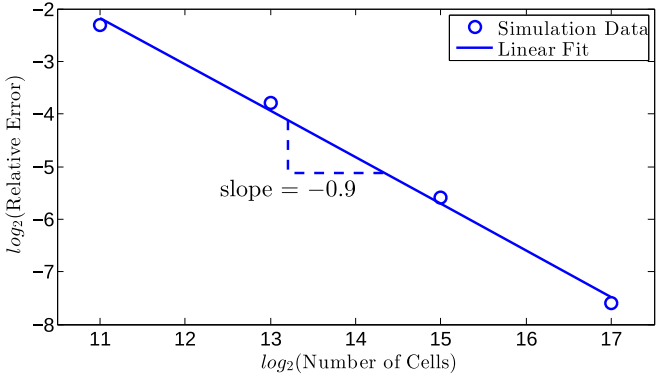}
	\caption{(Color online) Relative error of the critical Dean number as function of the number of grid points for a cylindrical channel with aspect ratio $\eta = r_i/r_o = 0.9$.}
	\label{fig:paper_size_study}
\end{figure}

To obtain an estimation of the resolution error in our simulations, we have measured the relative error of the critical Dean number for different resolutions. The relative error is defined as the relative deviation of the critical Dean number from the corresponding reference value given by Finlay \textit{et al.} in Ref. \cite{finlay1990}. Fig. \ref{fig:paper_size_study} depicts the dependence of the relative error on the grid resolution for a cylindrical channel with aspect ratio $\eta = r_i/r_o = 0.9$. As one can see, the relative error decreases rapidly when the grid resolution is increased. This shows that, within an error of $1\%$, our numerical results correspond to the physical values and are not affected by finite resolution effects.

\section{Flow through Double-Curved Channel}\label{sec:ellipsoid}

\begin{figure}
	\includegraphics[width=0.95\columnwidth]{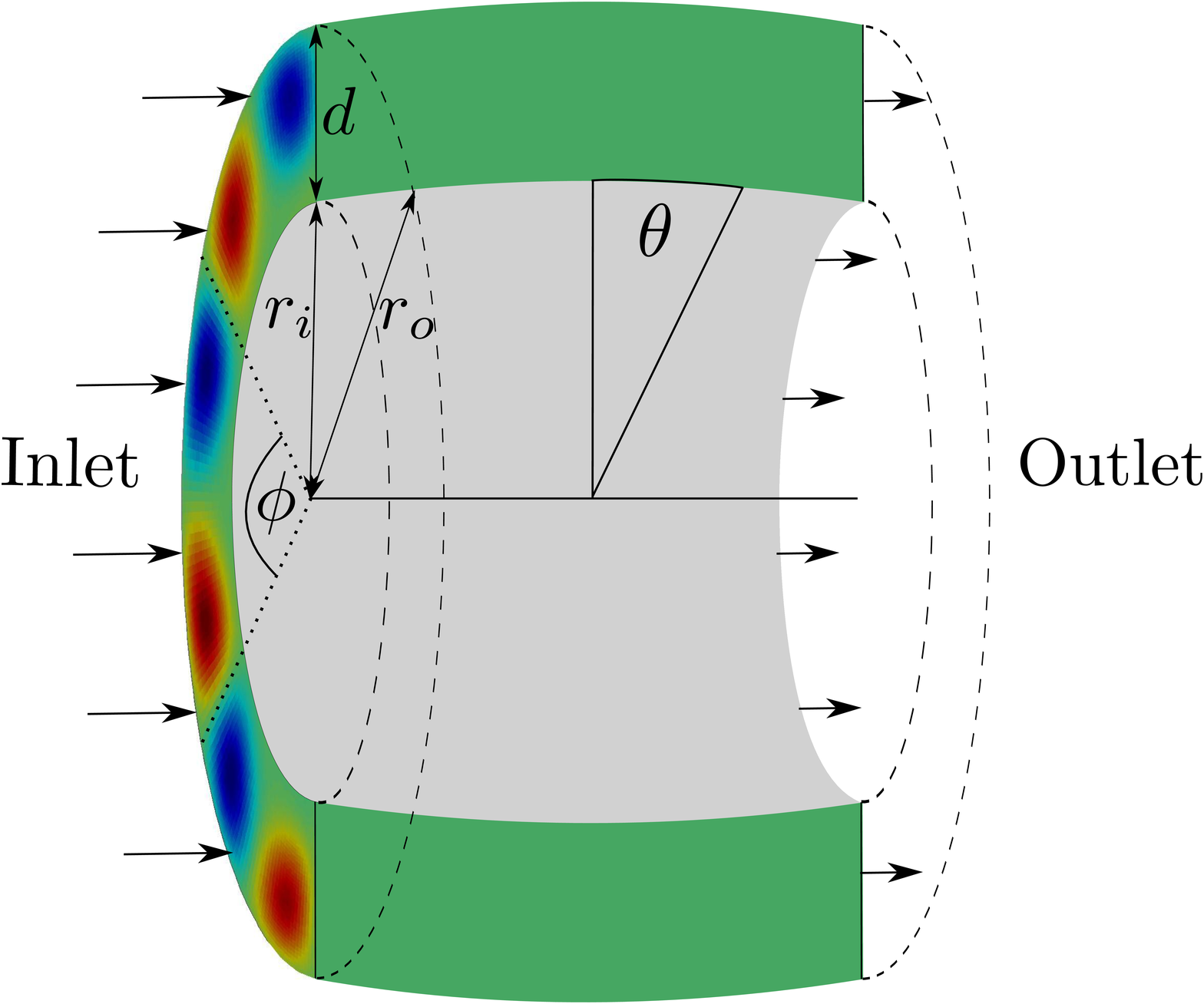}
	\caption{(Color online) Geometry of the double-curved channel. The colored spots on the channel cross-section at the inlet illustrate the strength of the flow vorticity. The dashed lines indicate the periodicity of the channel in $\ph$-direction.}
	\label{fig:ellipsoid_geometry}
\end{figure}

We now consider a more complex
geometry, namely a double-curved channel (see Fig. \ref{fig:ellipsoid_geometry}).
This geometry follows from the cylindrical configuration by introducing a
second curvature along the $z$-axis, which leads to a geometry that can best be described in terms of ellipsoidal coordinates $(x^1,x^2,x^3) = (r,\th,\ph)$, defined by
\begin{align*}
	x &= r \, a \, \cos \th \cos \ph, \\
	y &= r \, b \, \sin \th \cos \ph, \\
	z &= r \, c \, \sin \ph,	
\end{align*}
where $a,b,c$ are the lengths of the three semi-principle axes of the
ellipsoid. All vector fields and tensors are expressed in terms of the basis
$(\mathbf e_r, \mathbf e_\th, \mathbf e_\ph) = (\frac{\del}{\del r},
\frac{\del}{\del \th}, \frac{\del}{\del \ph})$.

Since finite-size effects in channels of
finite height are known to have a strong influence on the Dean
instability (as studied in Ref. \cite{finlay1990} for the case of two concentric
cylinders), we impose periodicity along the $\phi$-direction by choosing
$a=c=1$. This corresponds to the periodicity in $z$-direction for the
cylindrical case and avoids finite-height complications. The corresponding
metric components are given by 
\begin{align*}
	g_{rr} &= \sin^2 \ph + \cos^2 \ph \left( \cos^2 \th + b^2 \sin^2 \th \right),
	\\
	g_{\th\th} &=  r^2 \cos^2 \ph \left( \sin^2 \th + b^2 \cos^2 \th \right), \\
	g_{\ph\ph} &=  r^2 \left( \cos^2 \ph + \sin^2 \ph \left( \cos^2 \th + b^2 \sin^2 \th
	\right) \right), \\
	g_{r\th} &= g_{\th r} = r \left( b^2-1 \right) \cos^2 \ph \, \cos \th \, \sin
	\th,
	\\
	g_{r\ph} &= g_{\ph r} = r \left( 1 - b^2 \right) \sin^2 \th \, \cos \ph \,
	\sin \ph,  \\
	g_{\th\ph} &=  g_{\ph\th} = r^2 \left( 1 - b^2 \right) \cos \ph \, \sin \ph \,
	\cos \th \,	\sin \th,
\end{align*}
with non-vanishing Christoffel symbols
\begin{align*}
	\G^r_{\th\th} &= -r \cos^2 \ph, \\
	\G^r_{\ph\ph} &= -r, \\
	\G^\th_{r\th} &= \G^\th_{\th r} = \G^\ph_{r\ph} = \G^\ph_{\ph r}= 1/r, \\
	\G^\th_{\th\ph} &= \G^\th_{\ph\th} = -\tan \ph, \\
	\G^\ph_{\th\th} &= \sin \ph \, \cos \ph.
\end{align*}

In our simulations, we use a D3Q41 lattice of size $L_r \times L_\th
\times L_\ph = 64 \times 64 \times 128$. The lattice indices $(\hat r, \hat
\th, \hat \ph) \in [1,L_r] \times [1,L_\th] \times [1,L_\ph]$ are related to the
physical units by 
\begin{align*}
	r = r_i + \hat r \cdot \dr , \quad
	\th = \left( \hat \th - \frac{L_\th}{2} \right) \dth , \quad
	\ph = \left( \hat \ph - \frac{L_\ph}{2} \right) \dph ,	
\end{align*}
and, as in the cylindrical case, we choose $\dr = \dth = \dph = \dt = (1 - \eta)/(\eta L_r)$, where $\eta = r_i/r_o$ denotes the aspect ratio of the channel.

The inner and outer walls of the channel at $r = r_i$ and $r = r_o$ are
implemented in the same way as for the cylindrical case by using Dirichlet boundary conditions for
the fluid velocity, $u(r_o) = u(r_i) = 0$. We further use open boundary
conditions in $\th$-direction as well as periodic boundaries in $\ph$. Since by construction the channel is periodic in $\phi$, we can restrict the simulation domain to a circular sector in $\phi$ (as indicated by the dotted lines in Fig. \ref{fig:ellipsoid_geometry}). Note that by restricting to a finite sector and considering periodic boundary conditions, a symmetry condition is imposed on the solution, which might exclude some phases in the bifurcation diagram. However, since this is also the case for the cylindrical channel to which we want to compare, we do not take those phases into account. In particular, the double-curved channel is constructed in such a way that it approaches the geometry of the cylindrical channel (Fig. \ref{fig:cylinder_geometry}) for increasing inner radii $r_i$.
The open channel boundaries at $\th = \pm \frac{L_\th}{2} \, \dth$ define an inlet and an outlet for the flow, as can be seen in Fig. \ref{fig:ellipsoid_geometry}. In order to drive the fluid through the channel, an external force in $\th$-direction can be applied, which for incompressible flow is physically equivalent to a pressure gradient between the inlet and the outlet.

We study the bifurcation from curved channel Poiseuille flow to Dean vortex flow  by varying the
Dean number, 
\begin{align}
	\De = \frac{\langle v_\theta \rangle d}{\nu} \cdot \sqrt{\frac{d}{\R_\th}},
\end{align}
where $d = r_o - r_i$ and $\nu = \left( \tau - 1/2 \right) c_s^2 \dt$. $\R_\th$ denotes the streamwise curvature radius of the inner wall, given by $\R_\th = r_i b^2$ at $\th = 0$. The mean azimuthal velocity
$\langle v_\theta \rangle$ is calculated as follows,
\begin{align*}
	\langle v_\theta \rangle = \frac 1 S \int_S u^{\th} \, r^2 \,b \cos \ph\, dr d\ph,
\end{align*}
where $S = \int r\, dr d\ph$ is the cross-sectional area of
the channel and $u^\th$ is the azimuthal component of the velocity field $u =
u^r \mathbf e_r + u^\th \mathbf e_\th + u^\ph \mathbf e_\ph$. Like in the
cylindrical case, we measure the average vorticity in the streamwise direction
at $\th = 0$, given by
\begin{align*}
	\langle \w^\th \rangle = \frac 1 S \int_S \left| \pdiff{u^r}{\ph} -
	\pdiff{u^\ph}{r}\right| \, r\, b \cos\ph\, dr d\ph.
\end{align*}

We study two different cases: In the first case, we vary the streamwise
curvature radius of the channel $\R_\th = r_i b^2$ by changing the length of the
semi-principal axes $b$ while the inner radius $r_i$ is kept fixed. 
In the second case, we study the effect of the cross-sectional
curvature radius $\R_\ph = r_i$ along the $\ph$ direction by varying the inner
radius $r_i$ while keeping the streamwise curvature radius $\R_\th$ fixed.

\subsection{Variation of streamwise curvature}

First, we vary the curvature radius along the flow direction, $\R_\th = r_i b^2$, by
varying $b$ at fixed cross-sectional curvature radius $\R_\ph = r_i = 1$. For the
channel, we choose an aspect ratio of $\eta = r_i/r_o = 0.9$. For all the simulations of the double-curved channel, we set the relaxation time $\tau$ to $1$ in order to work in the desired parameter range of the Reynolds number (and Dean number), keeping at the same time a good computational performance. The fluid is initialized with a uniform
mass distribution by setting $\rho = \rho'\sqrt g = 1$ at $t=0$. The Dean number
is varied by changing the strength of the driving force. 

\begin{figure}
	\includegraphics[width=\columnwidth]{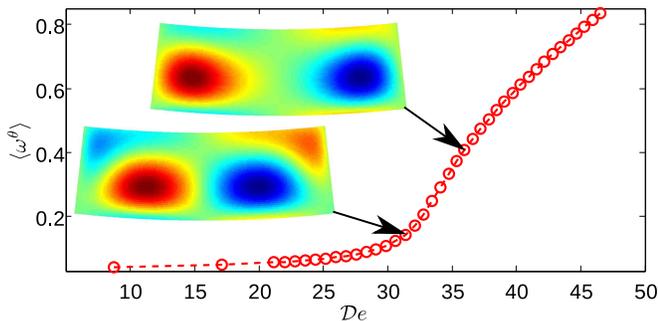}
	\caption{(Color online) Average vorticity $\langle \w^\th \rangle$ as function of the Dean number $\De$ for a double-curved channel with inner radius $r_i = 1$, aspect ratio $\eta = r_i/r_o = 0.9$ and $b = 0.9$. The bifurcation from Poiseuille flow to vortex flow occurs at $\De \approx 30.5$. The colored pictures depict the vorticity on a channel cross-section, where the blue (right) and red (left) spots correspond to clockwise and counterclockwise rotating vortices, respectively.}
	\label{fig:paper_ellipsoid_vorticity_plot}
\end{figure}

Again, we plot the average vorticity $\langle \w^\th \rangle$ as function of the Dean number $\De$ in order to determine the critical Dean number, at which the vorticity suddenly increases. Exemplarily, Fig. \ref{fig:paper_ellipsoid_vorticity_plot} shows the vorticity curve for $b = 0.9$. As can be observed, the vorticity begins to increase at Dean number $\De_c \approx 30$, which indicates an imperfect bifurcation from Poiseuille flow to 2-cell vortex flow. Compared to the cylindrical channel, this transition is however rather smooth. The colored pictures in Fig. \ref{fig:paper_ellipsoid_vorticity_plot} depict the vorticity $\w^\th$ on a cross-section. 

\begin{figure}
	\includegraphics[width=\columnwidth]{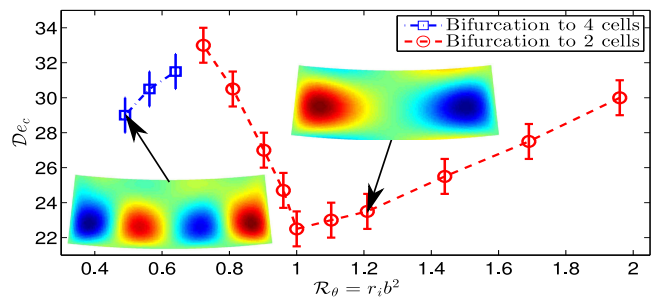}
	\caption{(Color online) Critical Dean number at the bifurcation points versus the streamwise curvature radius $\R_\th = r_i b^2$ for a double-curved channel with aspect ratio $\eta = r_i/r_o = 0.9$. The colored pictures depict the streamwise vorticity of the flow on a channel cross-section.}
	\label{fig:paper_critical_dean_ellipsoid}
\end{figure}

We have measured the critical Dean number at the bifurcation point for different streamwise curvatures $\R_\th = r_i b^2$. The results are depicted in Fig. \ref{fig:paper_critical_dean_ellipsoid}, where errorbars represent the uncertainty in reading off the critical Dean number from the vorticity curve. Two bifurcation points have been observed: a bifurcation from Poiseuille flow to 2-cell vortex flow for $\R_\th > 0.7$ and a bifurcation from Poiseuille flow to 4-cell vortex flow for $\R_\th < 0.7$. Fig. \ref{fig:paper_critical_dean_ellipsoid} shows that the critical Dean number for the bifurcation to 2-cell vortex flow possesses a minimum at a streamwise curvature $\R_\th = 1.00 \pm 0.05$, which corresponds to the spherical geometry, where all semi-principal axes of the ellipsoid are equal. For $\R_\th = 1.00 \pm 0.05$, the instability occurs already at a relatively low Dean number of $\De_c \approx 22.5$. 

For $\R_\th > 1$, the geometry of the channel is equivalent to a streamwise stretched ellipsoid, such that the streamwise curvature is expected to have only a minor influence on the instability relative to the spherical case $\R_\th = 1.00 \pm 0.05$. Still, the critical Dean number increases almost linearly with $\R_\th$ for $\R_\th > 1$, which suggests that the second curvature $\R_\ph$ dominates the instability in this regime.

For $\R_\th < 1$, on the other hand, the geometry of the channel resembles an ellipsoid compressed in streamwise direction, and the instability is dominated by the increasing streamwise curvature. For $\R_\th < 1$, we have observed the following behavior of the flow next to the bifurcation point: 
Right before the vortices begin to develop, there are four regions of slightly increased vorticity: Two next to the center of the outer wall at $r_o$ and two next to the corners of the inner wall at $r_i$. This can be seen in the lower colored picture in Fig. \ref{fig:paper_ellipsoid_vorticity_plot}, next to the bifurcation point. For $0{.}7 < \R_\th < 1$, the two inner regions dominate and finally form two major vortices (see the upper colored picture in Fig. \ref{fig:paper_ellipsoid_vorticity_plot}). For $\R_\th < 0{.}7$, on the other hand, the two vorticity regions in the corner also grow in strength and, together with the two inner regions, they finally form four major vortices (see the left colored picture in Fig. \ref{fig:paper_critical_dean_ellipsoid}). This behavior is different from the case $\R_\th > 1$, where we observe only two regions of increased vorticity in the beginning, which finally form two major vortices.

\subsection{Variation of cross-sectional curvature}

We also have studied the effect of the cross-sectional curvature radius $\R_\ph = r_i$ on the Dean instability by varying the inner radius $r_i$ of the channel. The
streamwise curvature radius $\R_\th = r_i b^2$ is set to $1$ and is kept fixed in
all simulations by setting $b = 1/\sqrt{r_i}$. The aspect ratio of the channel
is set to $\eta = 0.9$, and the lattice spacing in radial direction is given
by $\dr = (1-\eta)/(\eta L_r)$. In order to keep the physical dimensions of
the channel fixed when $r_i$ varies, the lattice spacings in $\th$-
and $\ph$-direction are rescaled accordingly: $\dth = \dr/(b\,r_i)$, $\dph = \dr/{r_i}$. 

\begin{figure}
	\includegraphics[width=\columnwidth]{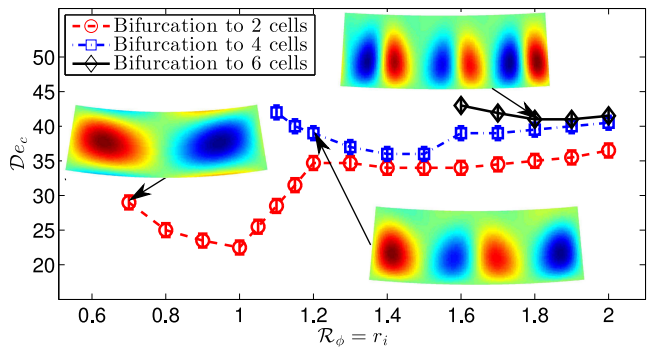}
	\caption{(Color online) Critical Dean number at the bifurcation points as function of the cross-sectional curvature radius $\R_\ph = r_i$. The colored pictures depict the streamwise vorticity of the flow on a channel cross-section.}
	\label{fig:paper_critical_dean_ellipsoid_vary_R0}
\end{figure}

Like in previous studies, we have analyzed the vorticity curve as function of the Dean number. Depending on the cross-sectional curvature radius $\R_\ph$, we have found three different bifurcations: a bifurcation from Poiseuille flow to 2-cell vortex flow, from 2 cells to 4 cells as well as from 4 cells to 6 cells. Fig. \ref{fig:paper_critical_dean_ellipsoid_vary_R0} depicts the dependence of the critical Dean number at the transition points on the curvature radius $\R_\ph$. The first bifurcation is the bifurcation from curved channel Poiseuille flow to 2-cell vortex flow. As can be observed, the critical Dean number for this transition possesses a minimum at $\R_\ph = 1.00 \pm 0.05$, corresponding to a spherical channel geometry, and the minimum value of the critical Dean number, $\De_c \approx 22.5$ at $\R_\ph = 1.00 \pm 0.05$ coincides with the corresponding value in Fig. \ref{fig:paper_critical_dean_ellipsoid} for the spherical case ($\R_\th = 1$). 
As can be seen in Fig. \ref{fig:paper_critical_dean_ellipsoid_vary_R0}, the critical Dean number increases with the cross-sectional curvature $1/\R_\ph$ for $\R_\ph < 1$. This suggests that for $\R_\ph < 1$ the instability is dominated by the cross-sectional curvature $1/\R_\ph$. For $\R_\ph > 1$ on the other hand, the critical Dean number grows almost linearly from $22.5$ at $\R_\ph=1$ to a value of about $35$ at $\R_\ph=1.2$, while a second bifurcation from 2-cell to 4-cell vortex flow begins to appear at higher Dean numbers. In this range, the instability is dominated by the perpendicular streamwise curvature $1/\R_\th$. For $\R_\ph > 1.2$, the critical Dean number of the first bifurcation stays more or less constant, whereas the threshold for the second bifurcation decreases further and further until it reaches a minimum at $\R_\ph = 1.55 \pm 0.05$. At this point, the third bifurcation from 4-cell to 6-cell vortex flow appears, while the critical Dean number at the first bifurcation point approaches the corresponding value for the cylindrical channel in the limit $\R_\ph \rightarrow \infty$. 

From the qualitative point of view, the three curves in Fig. \ref{fig:paper_critical_dean_ellipsoid_vary_R0} show a similar behaviour: Starting from different values of $\R_\ph$, within error bars all curves decrease to a minimum value right before the next bifurcation emerges. When the next bifurcation appears, the critical Dean number of the lower bifurcation increases to a rather constant value, which grows only slightly towards $\R_\ph \rightarrow \infty$.

\begin{figure}
	\includegraphics[width=\columnwidth]{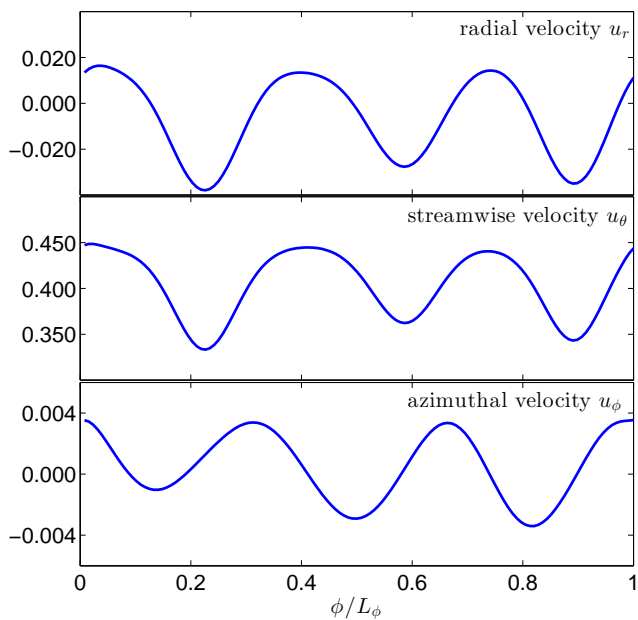}
	\caption{(Color online) Velocity profile of 6-cell vortex flow in a double-curved channel with aspect ratio $\eta = 0.9$, streamwise curvature $\R_\th = r_i b^2 = 1$ and cross-sectional curvature $\R_\ph = r_i = 1.8$. The corresponding Dean number is $\De = 44$.}
	\label{fig:paper_ellipsoid_velocity_plot}
\end{figure}

Physically, the appearance of 4-cells or 6-cells can be explained by the splitting mechanism described in Ref. \cite{guo1991splitting}: By increasing the Dean number, the Dean vortices can split up and form new vortex pairs. Although, one might wonder why there is no direct bifurcation from 4 cells to 8 cells, since in theory, the 4 vortices should be completely indistinguishable. In practice, however, the symmetry between the cells is broken by small perturbations, which leads to the splitting of only one vortex pair, resulting in the 6-cell solution. From the numerical point of view, we have checked the physicality of the 6-cell solution by doubling the grid resolution as well as by changing the triggering mechanism, finding the same result in both cases. Exemplarily, Fig. \ref{fig:paper_ellipsoid_velocity_plot} shows the cross-sectional velocity profile of a 6-cell vortex flow for $\R_\ph = 1.8$ and $\De = 44$.

\section{Conclusions}

Summarizing, we have studied the Dean instability in a double-curved channel, using our previously developed LB algorithm in general coordinates. The double-curved channel is characterized by a streamwise curvature as well as by a perpendicular cross-sectional curvature. In analogy with cylindrical channels, which have been widely studied in the past, we have observed a bifurcation from primary curved channel Poiseuille flow to secondary Dean vortex flow, which is characterized by counter-rotating vortices. In particular, we have measured the critical Dean number at the bifurcation points as function of the geometrical properties of the channel. 

At first, we have varied the streamwise curvature radius $\R_\th$ while keeping the cross-sectional curvature radius $\R_\ph$ fixed. We have found that the critical Dean number at the bifurcation from Poiseuille flow to 2-cell vortex flow is minimal for $\R_\th \approx \R_\ph$, where the channel possesses a spherical symmetry. For channels with weaker or stronger streamwise curvature, the critical Dean number increases almost linearly. For strongly-curved channels, we even have observed bifurcations from Poiseuille flow to 4-cell vortex flow. 

Secondly, we also have varied the cross-sectional curvature radius $\R_\ph$ while keeping the streamwise curvature fixed. Again, we have found that the lowest Dean number at which the Dean instability occurs corresponds to the spherically symmetric configuration, in which both curvature radii are approximately equal, $\R_\th \approx \R_\ph$, as before. When the cross-sectional curvature decreases towards the cylindrical channel limit, $\R_\ph \rightarrow \infty$, higher order bifurcations from 2-cell flow to 4-cell flow and even from 4-cell flow to 6-cell vortex flow come into play, while the critical Dean number of the first bifurcation from Poiseuille flow to 2-cell flow approaches the corresponding value for the cylindrical channel. 

\medskip

For the simulations, we have improved our previously developed LB algorithm in general coordinates to simulate flow in curved channels with complex geometries. The improved LB method has been validated for the case of flow through a cylindrical channel, for which we have measured the critical Dean number at the transition from laminar flow to Dean vortex flow for different aspect ratios of the channel. The results agree very well with numerically obtained results by Finlay \textit{et al.} \cite{finlay1990}. In addition to the linear stability analysis by Finlay \textit{et al.}, we have observed a second bifurcation from 2-cell to 4-cell vortex flow, supporting the existence of a second critical Dean number at the second bifurcation point.

The double-curved channel is implemented using an ellipsoidal coordinate system, which enters in our algorithm simply through the metric tensor. By using contravariant coordinates, the LB equation automatically adapts to the new geometry, which demonstrates the great advantage of our method when dealing with complex three-dimensional geometries. As we use generalized coordinates which are
perfectly adapted to the channel geometry, the channel boundaries can be implemented accurately without using the staircase approximation.

\medskip

\begin{acknowledgements}
  We acknowledge financial support from the European Research Council (ERC) Advanced
  Grant 319968-FlowCCS.
\end{acknowledgements}

\bibliography{citations.bib}

\end{document}